\documentclass[twocolumn,preprintnumbers,showpacs,showkeys,amsmath,amssymb,epsf]{revtex4}
\usepackage{graphicx}
\usepackage{dcolumn}
\usepackage{bm}
\begin{document}
\title{Particle current fluctuations in a particle-nonconserving process}
\author{Pegah Torkaman}
\author{Farhad H. Jafarpour}
\email{farhad@ipm.ir}
\affiliation{Physics Department, Bu-Ali Sina University, 65174-4161 Hamedan, Iran}
\date{\today}
\begin{abstract}
We have considered a one-dimensional coagulation-decoagulation system of classical particles on a finite lattice with reflecting boundaries.
It is known that the system undergoes a phase transition from a high-density to a low-density phase. Using a matrix product approach we 
have obtained an exact expression for the average entropy production rate of the system in the thermodynamic limit. We have also 
performed a large deviation analysis for fluctuations of entropy production rate and particle current. It turns out that
the characteristics of the kink in the large deviation function can be used to spot the phase transition point. We have found that for very weak driving 
field (when the system approaches to its equilibrium) and also for very strong driving field (when the system is in the low-density phase)
the large deviation function for fluctuations of entropy production rate is almost parabolic while in the high-density phase it prominently deviates 
from Gaussian behavior. The validity of the Gallavotti-Cohen fluctuation relation for the large deviation function for particle current is also verified. 
\end{abstract}
\pacs{05.40.-a,05.70.Ln,05.20.-y}
\keywords{entropy production, current fluctuations, large deviations in non-equilibrium systems, stochastic particle dynamics (theory)}
\maketitle
\section{Introduction}
Physical systems in nature are either in or out of equilibrium. One-dimensional out of equilibrium systems usually exhibit unique critical
behaviors such as phase transitions and shock formation\cite{Sch01}. These properties make them very interesting to study from both mathematical 
and physical point of views. Needless to say that many of these properties have not been fully understood yet.\\
An out of equilibrium system is usually exposed to a flux of matter or energy. This results in a non-zero probability current 
between different microstates of the system and also entropy production. It is known that for a system in equilibrium, where 
detailed balance is hold, the entropy production is deterministically zero. This is the reason why the entropy production can be an indicator for being out of equilibrium. \\
In recent years much attention has been paid to the study of fluctuation theorems in out of equilibrium 
systems~\cite{LebSpo,HarSch07,K07,Tou09}. It has been shown that for most of driven Markov processes the fluctuation theorem holds.
Symmetry properties of the large deviation functions of generalized current fluctuations have also been studied. 
These large deviation functions are also shown to obey a Gallavotti-Cohen (GC) type symmetry 
in systems with a finite state space. In contrast, for those systems with unbounded state space, such as the one-dimensional partially 
asymmetric zero-range process with open boundaries, the distribution of large current fluctuations does not satisfy the GC 
symmetry~\cite{HarRakSch05,HarRakSch06,RakHar08}. Other GC type symmetries have been found in a restricted class of 
Markov jump processes where the microscopic transitions have a particular structure and satisfy certain constraints~\cite{BarCheHinMuk12}. \\
In~\cite{SHT12} a large deviation analysis for fluctuations of partial and total particle currents in a zero-range process on a simple diamond 
lattice with open boundary conditions has been done. The validity of the GC fluctuation relation for these particle currents is investigated 
and it has been found that the fluctuation relation is not satisfied for partial particle currents between sites even if it is satisfied for the total 
particle current flowing between the boundaries.\\
So far, most of studies on the validity or breakdown of the GC symmetry and the fluctuation theorem 
have been mostly concentrated on boundary-driven systems. In this paper we consider an exactly 
solvable one-dimensional coagulation-decoagulation system. It is known that the system undergoes 
a phase transition in the steady state from a high-density to a low-density phase. 
We aim to study the large deviation functions for the fluctuations 
of entropy production rate and particle current and investigate their symmetries in long-time limit. \\
It turns out that the average entropy production rate of this system can be calculated exactly using a matrix product method. 
The average entropy production rate near the transition point changes abruptly, hence it 
can be used as a signal for the presence of a phase transition in the system. Studying the 
large deviation function for the entropy production rate reveals that the widely observed kink in 
the large deviation function disappears in the limit of large driving fields. The kink is also negligible 
as the driving field vanishes. However in the intermediate driving field regime the kink is prominent. 
This indicates that the characteristics of the kink in the large deviation function for the entropy production rate
can be used to investigate the phase transition in the system.\\ 
Our investigations also show that the large deviation functions for the local particle currents, defined as the net particle current 
through two consecutive lattice sites, do not satisfy the GC fluctuation relation. In contrast, the large 
deviation function for what we call the global particle current, defined as the sum of the local particle currents, 
satisfies the GC fluctuation relation. \\
This paper is organized as follows: in the second section we will define the process and summarize the known 
results about its steady state. The second section is dedicated to the mathematical basis of the fluctuation theorem.
Exact expression for average entropy production rate and its asymptotic behavior in the thermodynamic limit will be 
given in the fourth section. We will also perform a large deviation analysis for fluctuations of the entropy production 
in the fifth section. In the sixth section we will define the particle currents and check the validity of the GC fluctuation 
relation for these quantities. We will finally summarize the results.
\section{Definitions and known results}
Let us consider a system of interacting particles on a one-dimensional lattice of length $L$ with reflecting boundaries. 
The bulk of the system is assumed to evolve in time according to the following rules
\begin{equation}
\label{Rules}
\begin{array}{ll}
A+A \rightarrow \emptyset+A & \mbox{with rate} \; \; \omega_{24} \\
A+\emptyset \rightarrow \emptyset+A & \mbox{with rate} \; \; \omega_{23} \\
A+A \rightarrow A+\emptyset & \mbox{with rate} \; \; \omega_{34} \\
\emptyset+A \rightarrow A+\emptyset & \mbox{with rate} \; \; \omega_{32} \\
\emptyset+A \rightarrow A+A & \mbox{with rate} \; \; \omega_{42} \\
A+\emptyset \rightarrow A+A & \mbox{with rate} \; \; \omega_{43} \\
\end{array}
\end{equation}
in which $A$ and $\emptyset$ stand for the presence of a particle and a hole in each lattice site respectively. Note that 
there is no injection or extraction at the boundaries. This particle-nonconserving stochastic process has already been studied 
extensively in related literature~\cite{HKP96,HSK96,JM08,JA081,JA082}. The system has two different steady 
states: An empty lattice is a trivial steady state.
\begin{figure}
\includegraphics[width=80mm]{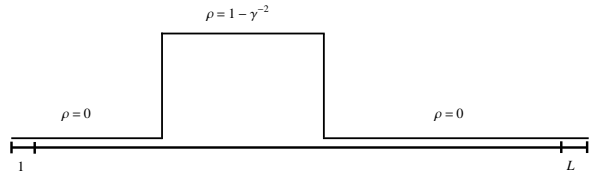}
\caption{\label{fig1} 
Simple sketch of a shock with two shock fronts on a lattice with $L$ lattice sites. 
The density of particles in each region is denoted by $\rho$.}
\end{figure}
It has been shown that the nontrivial steady state of this coagulation-decoagulation system is a matrix-product state (for a review of the 
matrix product approach see~\cite{BE}) provided that some constraints are fulfilled. More detailed investigations have shown 
that the steady state of the system can also be written in terms of a linear superposition of shock structures with two 
shock fronts (See FIG.~\ref{fig1} ). The shock fronts have simple random walk dynamics provided that some constraints on the microscopic 
reaction rates in~(\ref{Rules}) are met. These results confirm that the system has two different phases in the nontrivial steady 
state: a high-density (HD) phase and a low-density (LD) phase. In the HD phase the left (right) shock front moves preferentially to 
the left (right) while in the LD phase both shock fronts move preferentially to the left.\\
For the sake of simplicity we adopt the following choices for the microscopic reaction rates in~(\ref{Rules})~\cite{HKP96,HSK96}
\begin{equation}
\label{NewRules}
\begin{array}{l}
\omega_{24}=\omega_{23}=q^{-1},\\
\omega_{34}=\omega_{32}=q,\\
\omega_{42}=(\gamma^2-1) q,\\
\omega_{43}=(\gamma^2-1) q^{-1}.
\end{array}
\end{equation}
In terms of these newly defined microscopic reaction rates the system is in the HD phase for $q < \gamma$ while it is in the 
LD phase for $q > \gamma$. In the HD phase the bulk density of particles is equal to $\rho=1-\gamma^{-2}$ while it is equal 
to zero in the LD phase. On the coexistence line where $q=\gamma$ the bulk density of particles on the lattice changes linearly.\\
In the following section we will briefly review the mathematical basis of fluctuation theorem.
\section{Mathematical preliminaries}
Let us consider a continuous-time Markov process with a configuration space which is denoted by $\Omega$. We assume 
that a spontaneous transition from configuration $c$ to configuration $c'$ takes place with a rate $w_{c\to c'}$ where both 
$c$ and $c'$ belong to $\Omega$. The time evolution of the probability distribution $P(c,t)$, for the system being in $c$ at 
time $t$, is given by a master equation which can be written as
\begin{equation}
\label{ME}
\frac{d}{dt}P(c,t)= -\sum_{c'} H_{cc'}P(c',t)\, ,
\end{equation}
where $H$ is the Markov generator with elements
\begin{equation}	
H_{cc'}=\left\{
\begin{array}{ll} 
- w_{c'\to c} & \quad \textrm{ if } c\neq c'\\ \\
\sum_{c'\neq c} w_{c\to c'} & \quad \textrm{ if }  c=c' 
\end{array}\right.\,.
\end{equation}
Using quantum Hamiltonian formalism~\cite{Sch01} the master equation~(\ref{ME}) can be rewritten as 
\begin{equation}
\frac{\partial}{\partial t} \vert P(t) \rangle = - H \vert P(t) \rangle\, ,
\end{equation}
which is similar to the Schr\"odinger equation in imaginary time.\\
The generating function for any current $\mathcal{J}(t)$, which is a functional of 
the stochastic trajectory in configuration space $\Omega$, can be written as
\begin{equation}
\label{GF}
\langle e^{-\mu \mathcal{J}(t)} \rangle = \langle s\vert e^{-\hat{H}(\mu)t} \vert P(0) \rangle\, ,
\end{equation}
in which $\langle s \vert$ is a row vector with components $(1,1,1,\ldots)$ and that 
$\vert P(0) \rangle$ is the probability distribution vector at $t=0$. If the current $\mathcal{J}(t)$
changes its value by $\theta_{c \to c'}$ whenever a jump from $c \rightarrow c'$ occurs, 
then the matrix elements of the modified generator  $\hat{H}(\mu)$ in~(\ref{GF}) can be written as~\cite{HarSch07}
\begin{equation}	
\label{ModGen}
\hat{H}(\mu)_{cc'}=\left\{\begin{array}{ll} 
 -w_{c'\to c}\exp(-\mu \,\theta_{c'\to c}) & \quad \textrm{if } c\neq c'\\ \\
\sum_{c'\neq c} w_{c\to c'}  & \quad \textrm{if } c=c'
\end{array}\right.\,.
\end{equation}
In long-time limit the generating function~(\ref{GF}) can be written as
\begin{equation}
\lim_{t \to\infty} \langle e^{-\mu \mathcal{J}(t)} \rangle = e^{- t e(\mu)}\, ,
\end{equation}
in which $e(\mu)$ is given by the lowest eigenvalue of the modified generator~(\ref{ModGen}). 
Finally the Legendre transformation of $e(\mu)$, according to the Gr\"atner-Ellis theorem, gives 
the large deviation function~\cite{Tou09}
\begin{equation}
\hat{e}(x)=\textrm{max}_\mu(e(\mu)-x\mu)\,.
\end{equation}
According to the large deviation principle the probability distribution function, defined as the probability 
to observe a time-averaged value $x\equiv \mathcal{J}(t)/t$ of 
the current $\mathcal{J}(t)$ over time interval $[0,t]$, can now be written as 
\begin{equation}
\lim_{t\to\infty} \mathcal{P}(x,t)=e^{-t \hat{e}(x)}.
\end{equation}
In limit of large time the probability distribution function $\mathcal{P}(x,t)$ satisfies
\begin{equation}
\label{FT}
\frac{\mathcal{P}(-x,t)}{\mathcal{P}(x,t)}=e^{-Ext}
\end{equation}
where $E$ is a field conjugated to the flux $\mathcal{J}$. The relation~(\ref{FT}) can also be written as
\begin{equation}
\hat{e}(-x)-\hat{e}(x)=Ex
\end{equation}
which is also known as the CG fluctuation relation.
Depending on the definition of the time-integrated current $\mathcal{J}(t)$, the parameter $x$ can be regarded as entropy production rate or particle current. \\
Whenever the system jumps from $c$ to $c'$ the entropy in the environment changes by $\ln \frac{w_{c \to c'}}{w_{c' \to c}}$~\cite{S05}. 
In order to calculate the total entropy changes in a trajectory one takes $\theta_{c \to c'}=\ln \frac{w_{c \to c'}}{w_{c' \to c}}$ in~(\ref{ModGen}). 
The lowest eigenvalue of the modified generator~(\ref{ModGen}) satisfies what we know as the GC symmetry
\begin{equation} 
\label{GCE1}
e_{s}(\mu)=e_{s}(1-\mu)\, .
\end{equation} 
The large deviation function also satisfies the GC fluctuation relation
\begin{equation}
\label{GCE2} 
\hat{e}(-\sigma)-\hat{e}(\sigma)=\langle \dot{S}_{env}\rangle \sigma\, ,
\end{equation}
in which we have defined $\sigma:=S/\langle \dot{S}_{env}\rangle t$. 
The average entropy production rate in the environment can be calculated using
\begin{equation}
\label{AverageEntropyProductionD} 
\langle \dot{S}_{env}\rangle=\frac{d\,e_{s}(\mu)}{d\, \mu} \Big\vert_{\mu=0} \, .
\end{equation}
In order to analyze fluctuations of particle current $J$ in the steady state we need to count the net particle jumps  
during the time interval $t$. Whenever the system changes its configuration from $c$ to $c'$ a particle might 
contribute to particle current. To count total particle jumps in a trajectory one can take $\theta_{c \to c'}=\pm 1$ 
in~(\ref{ModGen}) for those configuration changes which contribute to particle current on the lattice in two different directions~\cite{HarSch07}.
In this case the lowest eigenvalue of the modified generator~(\ref{ModGen}) satisfies the GC symmetry
\begin{equation} 
\label{GCJ1}
e_{J}(\mu)=e_{J}(E-\mu)\, ,
\end{equation} 
where $E$ is the conjugate field. On the other hand, the large deviation function $\hat{e}(J)$ satisfies the GC fluctuation relation
\begin{equation}
\label{GCJ2} 
\hat{e}(-J)-\hat{e}(J)=EJ\, .
\end{equation}
\section{Average entropy production rate}
If $P^{\ast}(c)$ is the probability of being in configuration $c$ in the steady state, then besides the 
formula~(\ref{AverageEntropyProductionD}) the average entropy production in the environment is given by~\cite{S05}
\begin{equation}
\label{AverageEntropyProduction}
\langle \dot{S}_{env} \rangle= \sum_{c,c'}P^{\ast}(c)w_{c\to c'}\ln\frac{w_{c\to c'}}{w_{c'\to c}}\, ,
\end{equation}
provided that all transitions are reversible. Recent investigations show that the 
behavior of the average entropy production in systems with out of equilibrium 
phase transitions changes at the critical point~\cite{Gas04,CT05,ACR10,TO12,O12,BH12};
therefore, it plays an important role in classifying different nonequilibrium 
phase transitions.\\
In what follows we first show that~(\ref{AverageEntropyProduction}) can be calculated 
exactly for the system defined by~(\ref{Rules}) and~(\ref{NewRules}). In order to calculate the steady 
state probability distribution $P^{\ast}(c)$, one can use a matrix product 
approach. According to this approach $P^{\ast}(c)$ is written as a product of noncommuting operators 
which satisfy an algebra~\cite{BE}. For the system defined by~(\ref{Rules}) and~(\ref{NewRules}) it has been shown 
that there exists a four-dimensional matrix representation for the quadratic algebra of the system~\cite{HSK96,JA082}.
\begin{figure*}
\includegraphics[scale=0.85]{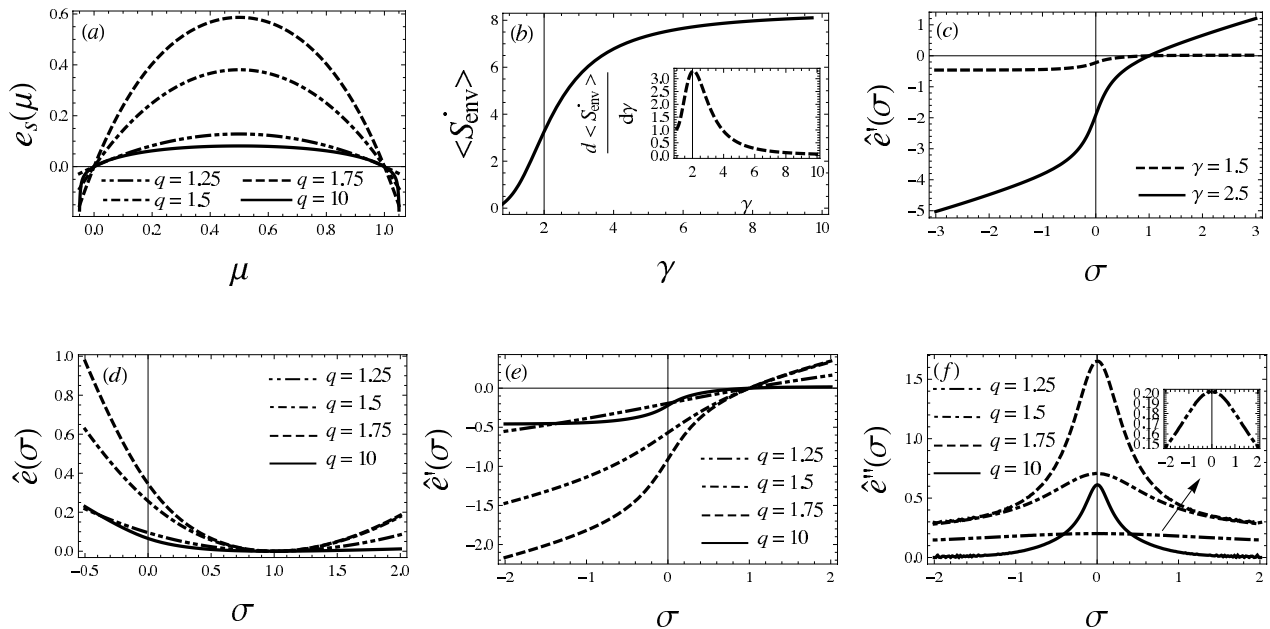}
\caption{\label{fig2} Numerical results for a system of length $L=4$: 
(a) The lowest eigenvalue of the modified generator for 
the entropy production for $\gamma=2$ and different values of $q$.  
(b) The average entropy production rate and its derivative for $q=2$ as 
a function of $\gamma$. The inset shows its first derivative respect to $\gamma$. 
(c) The first derivative of the large deviation function for the entropy production for $q=2$ 
and two different values of $\gamma$. In (d), (e) and (f) the large deviation function and 
its first and second derivatives are plotted for different values of $q$ and $\gamma=2$.}
\end{figure*}
Using the four-dimensional matrix representation and the results obtained in~\cite{HSK96,JA082} we find that in large-$L$ limit
the average entropy production rate in the environment is given by
\begin{equation}
\label{ExactAEP}
\langle \dot{S}_{env} \rangle  \;=\; 
\left\{
\begin{array}{ll}
\frac{2(q^2-1)(1-3\gamma^2+2\gamma^4) \ln q}{q \gamma^4}L& \mbox{for} \; q<\gamma, \\[4mm] 
\frac{2(\gamma^2-1)(q^2\gamma^2-1)(q^2+2\gamma^2(\gamma^2-1))\ln q}{(1+q^2)(q^2-\gamma^2)q\gamma^4} & \mbox{for} \; q>\gamma.
\end{array}
\right.
\end{equation}
As can be seen $\langle \dot{S}_{env} \rangle$ changes discontinuously at the transition point. 
In other words, the average entropy production rate per lattice site in the steady state $\langle \dot{S}_{env} \rangle/L$  
is a constant in the HD phase while it is zero in the LD phase. This can be easily explained as follows: in the HD
phase the lattice is nearly full of particles. This can be realized by the fact that in this phase the steady state is a linear
superposition of product shock measures with two shock fronts in which left shock front moves preferentially to the left 
while the right shock front moves preferentially to the right (see FIG.~\ref{fig1}). The more particles contribute into 
the reactions, the more entropy is produced in the environment. Since nearly all lattice sites contribute in entropy 
production, the average entropy production~(\ref{AverageEntropyProduction}) in the HD phase is 
proportional to the system size. In contrast, in the LD phase both shock fronts move preferentially  to the 
left (see FIG.~\ref{fig1}); therefore, the lattice is almost empty. In this phase much less particles contribute in entropy 
production. That is why the average entropy production per lattice site is zero in large-$L$ limit. 
We should also note that the average entropy production in the steady state~(\ref{AverageEntropyProduction}) 
is equal to zero for $q=1$. This is the value of $q$ for which the system is in equilibrium. We will discuss this later in forthcoming chapters. 
\section{Entropy fluctuations}
Finding an exact expression for the lowest eigenvalue of the modified generator for the entropy production $e_{s}(\mu)$ of a 
system of length $L$ is a formidable task; however, this can be done numerically for small lattices. 
Numerically exact results obtained for a system of length $L=4$ are given in FIG.~\ref{fig2}. We have 
plotted $e_{s}(\mu)$ as a function of $\mu$ for $\gamma=2$ in FIG.~\ref{fig2}(a). As $q \to 1$ the system
approaches to its equilibrium state, hence $e_{s}(\mu)$ is almost parabolic~\cite{DP11,MSS08}. It can be seen that $e_{s}(\mu)$
behaves almost the same way for large values of $q$. \\
In FIG.~\ref{fig2}(b) we have plotted the average entropy production rate obtained using~(\ref{AverageEntropyProductionD})
as a function of $\gamma$ for $q=2$. The inset in this figure shows a peak in the derivative of the average entropy production 
rate which becomes more prominent by increasing system size $L$ as~(\ref{ExactAEP}) has already predicted. 
This indicates that the behavior of $\langle \dot{S}_{env} \rangle$ can be used to spot the transition point. \\
In FIG.~\ref{fig2}(c) the first derivative of large deviation function for the entropy production rate respect to $\sigma$
for $q=2$ and two values of $\gamma$, one above and one below the transition point, is plotted. Using~(\ref{GCE2})
one can easily find that~\cite{DP11}
\begin{equation}
\label{Jump}
\hat{e}'(\sigma)\Big\vert_{\sigma_{0}}-
\hat{e}'(\sigma)\Big\vert_{-\sigma_{0}}=
2\hat{e}'(\sigma)\Big\vert_{\sigma_{0}}+\langle \dot{S}_{env} \rangle.
\end{equation}
This shows that the sudden jump in the first derivative of the large deviation function is more pronounced for large values of the 
average entropy production rate i.e. in the HD phase. One should recall that the average entropy production rate as a function of $\gamma$ 
is of order $L$ for $q<\gamma$ while it is of order of unity for $q>\gamma$. \\
In FIG.~(\ref{fig2})(d) we have plotted $\hat{e}(\sigma)$ for $\gamma=2$ and different values of $q$ . 
It is known that the large deviation function for the entropy production rate exhibits a kink at $\sigma=0$ which is a generic feature of 
the large deviation function and follows from the fluctuation theorem~\cite{DP11,MSS08}. As $q\to1$ the kink at $\sigma=0$ 
disappears and the curve becomes a parabola which is, as we mentioned, an indication for the system getting close to its 
equilibrium. It can be seen that $\hat{e}(\sigma)$ has almost the same behavior for $q >> 1$. As we will see in the next section the driving
field which drives the system out of equilibrium depend only on $q$.\\
In FIG.~\ref{fig2}(e) and FIG.~\ref{fig2}(f) we have plotted the first and second derivatives of the large deviation 
function for the entropy production rate for different values of $q$. The existence of a kink is best illustrated by these 
derivatives at $\sigma=0$. It can be seen that the jump in the first derivative of $\hat{e}(\sigma)$ at $\sigma=0$ 
disappears and its second derivative becomes a constant as $q\to1$ i.e. the system approaches an equilibrium steady state.
The inset in~FIG.~\ref{fig2}(f) shows that as long as $q \ne 1$ the system is out of equilibrium. 
While being in the HD phase $1 < q < \gamma$ the jump in the first derivative of the large deviation 
function increases as $q$ increases. Note that $\langle \dot{S}_{env} \rangle$ in the HD phase
is an increasing function of $q$ and of order of the system length $L$ (see~(\ref{ExactAEP})) and that the jump in the first derivative of the large deviation function 
is governed by $\langle \dot{S}_{env} \rangle$ as can be seen in~(\ref{Jump}). \\
In the LD phase $q > \gamma$ the average entropy production rate is negligible (of order of unity as can be seen in~(\ref{ExactAEP})) and one
expects that the jump in the first derivative of the large deviation function decreases in comparison to its value in the HD phase. This can also be seen
in FIG.~\ref{fig2}(e) and FIG.~\ref{fig2}(f) for $q >> 1$. We expect that this will be more prominent as the system size is increased. We conclude this section 
by noting that the characteristics of the kink (the jump in first derivative of the large deviation function) can be used as a criteria for spotting the phase transition point.\\
In the next section we will perform a large deviation analysis for the particle current in the system.
\section{Particle current fluctuations}
For the system defined by~(\ref{Rules}) and~(\ref{NewRules}) the particle density is not 
conserved. The time evolution of the average local particle density 
$\langle \rho_{k} \rangle (t)$ at a lattice site $k$ ($k=1,\cdots,L$) is given by
\begin{equation}
\label{PCEofM}
\frac{d}{dt}\langle \rho_{k} \rangle (t) =\langle J_{k-1} \rangle (t)- \langle J_{k} \rangle (t)+S_{k}(t)
\end{equation} 
where $\langle J_{k} \rangle (t)$ is called the average local particle current from lattice site 
$k$ to $k+1$ and $S_{k}$ is a source term. 
\begin{figure}
\includegraphics[width=85mm]{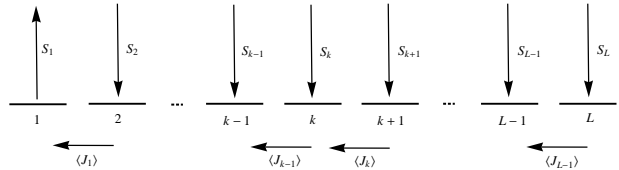}
\caption{\label{fig3} Local particle currents and external sources in a system of length $L$.}
\end{figure}
In the steady state the time dependency of the quantities will be dropped; therefore, the 
l.h.s of the equation~(\ref{PCEofM}) becomes equal to zero and we find (see FIG.~\ref{fig3})
\begin{equation}
\label{SSPCEofM}
S_{k}=\langle J_{k} \rangle-\langle J_{k-1}\rangle \;\; \mbox{for} \;\; k=1,\cdots,L.
\end{equation}
in which the average local particle current is defined as
\begin{eqnarray}
\langle J_{k} \rangle & = & \Big ( q \langle \rho_{k} \rho_{k+1} \rangle   - q^{-1} \langle \rho_{k} \rho_{k+1} \rangle \nonumber\\
&& +\Delta q \langle (1-\rho_{k}) \rho_{k+1} \rangle+  q \langle (1-\rho_{k}) \rho_{k+1} \rangle \nonumber \\
&& -\Delta q^{-1} \langle \rho_{k} (1- \rho_{k+1}) \rangle - q^{-1} \langle \rho_{k} (1-\rho_{k+1}) \rangle \Big ) \nonumber \\
& = & - q^{-1} (1+\Delta) \langle \rho_{k} \rangle+q(1+\Delta)\langle \rho_{k+1} \rangle  \nonumber \\
& & -\Delta(q-q^{-1})\langle \rho_{k} \rho_{k+1} \rangle.\nonumber
\end{eqnarray}
It is easy to verify that $\sum_{k=1}^{L}S_{k}=0$. We will also define an average global particle current as 
\begin{equation}
\label{GC}
\langle J \rangle =\sum_{k=1}^{L-1} \langle J_{k} \rangle.
\end{equation}
Using the matrix product approach one can easily calculate the average local particle currents in the 
steady state. It turns out that the exact expression for the average local particle current is given by
\begin{eqnarray}
\label{localcurrents}
\langle J_{k} \rangle & = & \frac{\left(1-\gamma ^2\right) \left(1-q^2\right) \gamma ^{-2 k+2 L-4} q^{-4 k+2 L-3}}
{\gamma ^{2 L}+(\gamma  q)^{2 L} \left(q^{2 L}-\gamma ^{2 L}\right)-q^{2 L}}\nonumber \\
& \times & \Big (\gamma ^{2 k} q^{2 L} (\gamma ^2 (\gamma ^2+q^2-1) (\gamma  q)^{2 k} \nonumber \\
& + & \gamma ^2(1-\gamma ^2)(1+q^4)-q^2) \nonumber \\
& + & q^{2 k} \gamma ^{2 L}(q^2(1-2 \gamma ^2) (\gamma  q)^{2 k} \nonumber \\
& + & \gamma ^2 q^2 (\gamma ^2 q^2-q^2+1)) \Big)
\end{eqnarray}
for $k=1,\cdots,L-1$. 
\begin{figure}
\includegraphics[scale=0.07]{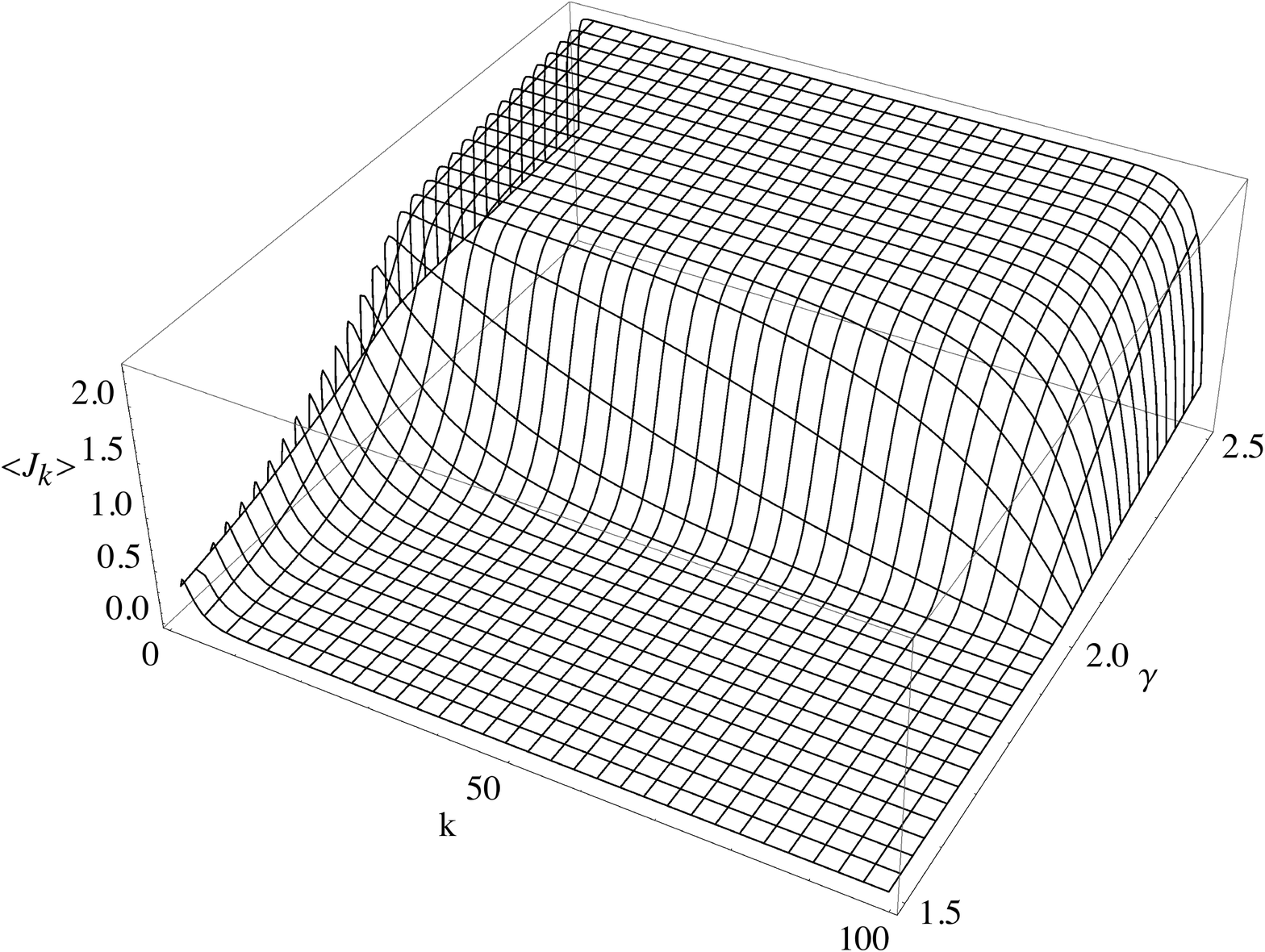}
\caption{\label{fig4} The average local particle currents~(\ref{localcurrents}) as a function of $\gamma$ for $q=2$ and $L=100$. }
\end{figure}
In FIG.~\ref{fig4} we have plotted~(\ref{localcurrents}) as a function of $\gamma$ for $q=2$ and $L=100$. It can be seen 
that for $\gamma  <  2$ i.e. in the LD phase, the average local particle currents are zero except in the vicinity of the left boundary. 
However, for $\gamma > 2$  i.e. in the HD phase, the average local particle currents are nonzero throughout the lattice. \\ 
In order to calculate the average local particle currents one can equivalently construct a modified generator for a jump process which counts the number 
of local or global particle jumps on a trajectory over a time interval $[0,t]$. This can be done using~(\ref{ModGen}) and 
an appropriate choice for $\theta_{c\to c'}$ as we explained in mathematical preliminaries section. The first derivative 
of the lowest eigenvalue of this operator respect to $\mu$ at $\mu=0$ gives the average particle current. \\
We have found that only the global particle current $J$ satisfies GC fluctuation relation~(\ref{GCJ2}) and 
that its average is proportional to the average entropy production rate
\begin{equation}
\label{SJ}
\langle \dot{S}_{env} \rangle=E \langle J \rangle
\end{equation}
where $E$ is equal to $\ln q^{2}$. The thermodynamic force $E$ which is conjugated to the flux of particle $\langle J \rangle$ 
physically corresponds to the effective driving field pushing particles.
\begin{figure*}
\centering
\includegraphics[scale=0.8]{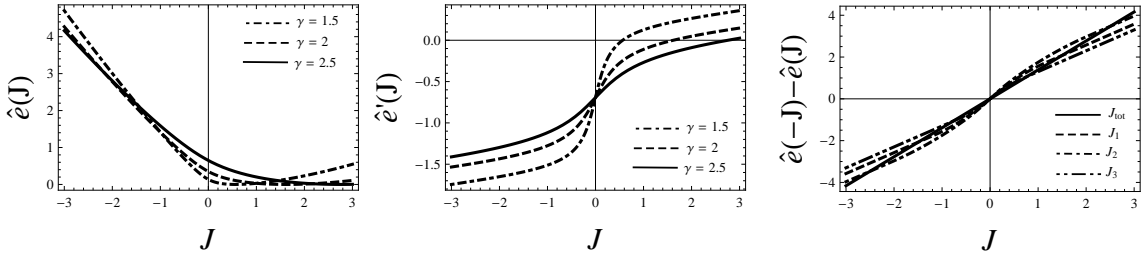}
\caption{\label{fig5} (Left to right) The large deviation function for the global particle 
current and its first derivative in a system of length $L=4$ for $q=2$ and three values 
of $\gamma$. These points are located below, above and at the transition point. The last figure shows 
that the large deviation function for the global particle current satisfies the GC 
fluctuation relation theorem~(\ref{GCJ2}).}
\end{figure*}
Our numerically verified results show that the lowest eigenvalue of the modified
generator for the global particle current and that of the entropy production satisfy 
\begin{equation}
e_{s}(\mu)=e_{J}(E\mu).
\end{equation} 
The fact that the local particle currents do not fulfill GC fluctuation relation has already been observed in a zero-range process on a 
diamond lattice with open boundaries\cite{SHT12}. In this paper we study an exactly solvable system with a particle-nonconserving dynamics. \\
Using the fact that the system is at equilibrium when $q=1$ we can calculate the conjugate field $E$ as follows. 
We imagine that there exists an equilibrium system with rates $w^\text{eq}_{c \to c'}$ which obey detailed balance. 
Now by applying an external field $E$, we recover our system defined in~(\ref{Rules}) 
and~(\ref{NewRules}) whose microscopic transition rates satisfy the following relation 
\begin{equation}
\label{Newrates}
w_{c\to c'}=w^\text{eq}_{c \to c'} e^{\frac{E}{2} \theta_{c\to c'}}
\end{equation} 
in which $\theta_{c\to c'}=1$ ($\theta_{c\to c'}=-1$) if the transition from $c$ to $c'$ is associated with a particle jump to the left 
(right) on the lattice. The relation~(\ref{Newrates}) gives $E=\ln q^2$ provided that the microscopic 
rates of the system in equilibrium are given by~(\ref{NewRules}) with $q=1$. \\
The matrix product approach predicts that the equilibrium probability distribution function $P^{\ast}_{eq}(c)$, i.e. the probability 
distribution function $P^{\ast}(c)$ at $q=1$, has the following properties
\begin{eqnarray}
&& P^{\ast}_{eq}(\cdots10\cdots)=P^{\ast}_{eq}(\cdots01\cdots),\nonumber \\
&& P^{\ast}_{eq}(\cdots11\cdots)=(\gamma^2-1)P^{\ast}_{eq}(\cdots01\cdots).
\end{eqnarray}
It is not hard to verify that the equilibrium probability distribution function satisfies the local detailed balance condition
given by~\cite{HarSch07}
\begin{equation}
w_{c \to c'} e^{-\frac{E}{2}\theta_{c \to c'}} P^{\ast}_{eq}(c)= w_{c' \to c} e^{-\frac{E}{2}\theta_{c' \to c}} P^{\ast}_{eq}(c').
\end{equation}
It is known that the local detailed balance condition leads to the GC symmetry of the global particle current~\cite{LebSpo}.\\ 
Let us now have a look at the behavior of $\langle J \rangle$ as a function of $q$. 
As we mentioned above, the driving force $E$ is zero at $q=1$ and therefore the system is at equilibrium. 
At this point $\langle J \rangle$ is zero. As $q$ is increased the driving force increases and hence
the system is driven out of equilibrium. The average global particle current is an increasing function of $q$ 
for $1 < q < \gamma$ i.e. in the HD phase. The phase transition occurs at $q=\gamma$. 
Above the transition point i.e. in the LD phase, $\langle J \rangle$ becomes negligible. \\
In FIG.~\ref{fig5} we have plotted the large deviation function for the global particle current and its first 
derivative for $q=2$ and three values of $\gamma$ in a system of length $L=4$. These point are chosen to be
above, below and at the transition point. As can be seen, a sudden jump exists in the first derivative of the large 
deviation function at $J=0$. It can be seen that in the LD phase the minimum of the large deviation function occurs 
at a point (which gives the average global particle current) close to zero while it is nonzero in the HD phase. 
This is in accordance with the average entropy production rate behavior as we explained before. \\ 
In order to check the validity of the GC fluctuation relation for the global and local particle currents
we have plotted~(\ref{GCJ2}) both for $J$ and $J_{k}$s in a system of length $L=4$ in FIG.~\ref{fig5}. It can be 
seen that only the global particle current satisfies the GC fluctuation relation in accordance with~(\ref{GCJ2}). 
The slop of this line is equal to $E$. For the local particle currents a linear behavior can be seen only in the vicinity of the origin. 
\section{Conclusion}
In this paper we have considered a one-dimensional classical system with reflecting boundaries and a particle-nonconserving dynamics. 
It is known that by varying the microscopic reaction rates, the system undergoes a phase transition from a LD phase to a HD phase. \\
Using a matrix product approach we have obtained exact expression for the average entropy production rate in the environment 
in the long time limit. It turns out that the average entropy production rate changes discontinuously at the phase transition point while 
it is zero at $q=1$ where the system is in equilibrium. \\
We have studied the entropy fluctuations in the system for $L=4$. We have found that the large deviation function for the entropy 
production rate becomes a parabola as the system approaches to its equilibrium. At zero entropy production rate a kink is observed 
in the large deviation function for the entropy production rate. The kink disappears at both very large and very small driving fields.
We expect that in large-$L$ limit the kink (also the discontinuity in the first derivative of the large deviation function 
for the entropy production rate) become more noticeable in the HD phase. \\
We have also investigated the validity of the GC fluctuation relation for the particle current in this system. We have considered two types of 
particle currents: local particle currents which are defined as particle currents between consecutive lattice sites and global particle current 
as a sum of these local particle currents. The average particle currents are calculated exactly. Our numerical investigations reveal 
that only the global particle current fulfills the GC fluctuation relation. Moreover, our analytical and numerical investigations show that the average
global particle current $\langle J \rangle$ is proportional to the average entropy production rate $\langle \dot{S}_{env} \rangle$.\\
Previous investigations have shown that the first derivative of the average entropy production rate displays a peak, a discontinuity or a divergence at criticality. 
In this paper we have shown that the stationary average entropy production rate in our system, defined by~(\ref{Rules}) and~(\ref{NewRules}), 
changes discontinuously at the critical point. It seems that different nonequilibrium phase transitions can be classified using criticality 
of average entropy production rate at a transition point.  



\begin{thebibliography}{99}
%
 \bibitem{Sch01} G. M. Sch\"utz, \textit{Phase transitions and critical phenomena}, 2001, vol. 19 3, London: Academic
%
\bibitem{LebSpo} J. L. Lebowitz, H. Spohn, J. Stat. Phys. 95 333 (1999)
\bibitem{HarSch07} R. J. Harris, G. M. Sch\"utz, J. Stat. Mech. P07020 (2007)
\bibitem{K07}  J. Kurchan, J. Stat. Mech. P07005 (2007)
\bibitem{Tou09} H. Touchette, Phys. Rep. 478 1 (2009)
%
\bibitem{HarRakSch05} R. J. Harris, A. R\'akos, G. M. Sch\"utz, J. Stat. Mech., P08003 (2005)
\bibitem{HarRakSch06} R. J. Harris, A. R\'akos, G. M. Sch\"utz, Europhys. Lett, 75 227 (2006)
\bibitem{RakHar08} A. R\'akos, R. J. Harris, J. Stat. Mech., P05005 (2008)
%
\bibitem{BarCheHinMuk12} A. C. Barato, R. Chetrite, H. Hinrichsen, D. Mukamel, J. Stat. Phys., 146 294 (2012)
%
\bibitem{SHT12} R. Villavicencio-Sanchez, R. J. Harris and H. Touchette, J. Stat. Mech. P07007 (2012)
%
\bibitem{HKP96} H. Hinrichsen, K. Krebs and I. Peschel Z. Phys. B 100 105 (1996)
\bibitem{HSK96} H. Hinrichsen, S. Sandow and I. Peschel J. Phys. A: Math. Gen. A 29 2643 (1996)
\bibitem{JM08} F. H. Jafarpour and S. R. Masharian, Phys. Rev. E 77, 031115 (2008)
\bibitem{JA081} F. H. Jafarpour and A. Aghamohammadi, Phys. Rev. E 78, 041108 (2008)
\bibitem{JA082} F. H. Jafarpour and A. Aghamohammadi, J. Phys. A: Math. Theor. 41 365001 (2008)
%
\bibitem{BE} R. A. Blythe, M. R. Evans, J. Phys. A Math. Theor. 40 R333-R441(2007)
%
\bibitem{S05} U. Seifert, Phys. Rev. Lett. 95 040602 (2005)
%
\bibitem{Gas04} P. Gaspard, J. Chem. Phys. 120 8898 (2004)
\bibitem{CT05} L. Crochik and T. Tome, Phys. Rev. E 72 057103 (2005)
\bibitem{ACR10} B. Andrae, J. Cremer, T. Reichenbach and E. Frey, Phys. Rev. Lett. 104 218102 (2010)
\bibitem{TO12} T. Tome and M. J. de Oliveira, Phys. Rev. Lett. 108 020601 (2012)
\bibitem{O12} M. J. de Oliveira, J. Stat. Mech. P12012 (2012)
\bibitem{BH12} A. C. Barato and H. Hinrichsen, J. Phys. A: Math. Theor. 45 115005 (2012)
%
\bibitem{DP11} S. Dorosz and M. Pleimling, Phys. Rev. E 83, 031107 (2011)
\bibitem{MSS08} J. Mehl, T. Speck, and U. Seifert, Phys. Rev. E 78, 011123 (2008)
\end{thebibliography}
\end{document}